\documentclass[10pt,pra,twocolumn,superscriptaddress,showpacs,floatfix]{revtex4}

\usepackage{graphicx}
\usepackage{amssymb}
\usepackage{times}
\usepackage{amsmath}
\usepackage{amsthm}
\usepackage{epstopdf}
\usepackage{verbatim}
\usepackage{hyperref}

\input epsf.tex
\usepackage{graphicx}
\usepackage{amsthm}

\begin{document}

\title{Long-distance continuous-variable measurement-device-independent quantum key distribution with discrete modulation}

\author{Hong-Xin Ma}\affiliation
 {Zhengzhou Information Science and Technology Institute, Zhengzhou 450004, China
}\affiliation{Synergetic Innovation Center of Quantum Information and Quantum Physics, University of Science and Technology of China, Hefei, Anhui 230026, China}
\affiliation
 {State Key Laboratory of Advanced Optical Communication Systems and Networks and Center of Quantum Information Sensing and Processing, Shanghai Jiao Tong University, Shanghai 200240, China
}

\author{Peng Huang}\thanks{Corresponding author: huang.peng@sjtu.edu.cn}
\affiliation
 {State Key Laboratory of Advanced Optical Communication Systems and Networks and Center of Quantum Information Sensing and Processing, Shanghai Jiao Tong University, Shanghai 200240, China
}

\author{Dong-Yun Bai}\affiliation
 {State Key Laboratory of Advanced Optical Communication Systems and Networks and Center of Quantum Information Sensing and Processing, Shanghai Jiao Tong University, Shanghai 200240, China
}

\author{Tao Wang}\affiliation
 {State Key Laboratory of Advanced Optical Communication Systems and Networks and Center of Quantum Information Sensing and Processing, Shanghai Jiao Tong University, Shanghai 200240, China
}

\author{Shi-Yu Wang}\affiliation
 {State Key Laboratory of Advanced Optical Communication Systems and Networks and Center of Quantum Information Sensing and Processing, Shanghai Jiao Tong University, Shanghai 200240, China
}

\author{Wan-Su Bao}\affiliation
 {Zhengzhou Information Science and Technology Institute, Zhengzhou 450004, China
}\affiliation{Synergetic Innovation Center of Quantum Information and Quantum Physics, University of Science and Technology of China, Hefei, Anhui 230026, China}

\author{Gui-Hua Zeng}\affiliation
 {State Key Laboratory of Advanced Optical Communication Systems and Networks and Center of Quantum Information Sensing and Processing, Shanghai Jiao Tong University, Shanghai 200240, China
}

\date{\today}

\begin{abstract}
We propose a long-distance continuous-variable measurement-device-independent quantum key distribution (CV-MDI-QKD) protocol with discrete modulation. This kind of discrete-modulated schemes have good compatibility with efficient error correction code, which lead to higher reconciliation efficiency even at low signal-to-noise ratio (SNR).
Security analysis shows that the proposed protocol is secure against arbitrary collective attacks in the asymptotic limit with proper use of decoy states.
And with the using of discrete modulation, the proposed CV-MDI-QKD protocol has simpler implementation and outperform previous protocols in terms of achievable maximal transmission distance, which precisely solve the bottleneck of the original Gaussian-modulated CV-MDI-QKD protocol.
\end{abstract}


\pacs{03.67.Hk, 03.67.-a, 03.67.Dd}
\maketitle

\section{Introduction}\label{Intr}
Quantum key distribution (QKD) \cite{NG02} allows two distant authenticated users to establish secure key through untrusted quantum and classical channels.
QKD has two main categories: one is discrete-variable (DV)QKD protocols \cite{BB84,E91,B92}, and the alternative is continuous-variable (CV) QKD protocols \cite{TC99,GG02,GG03,CS12}.
Compared with DVQKD protocols (such as BB84 protocol), CVQKD protocols have unique potentials of being effectively compatible with existing optical communication systems and using lower cost light sources and detectors.
In addition, CVQKD allow users to approximate the ultimate limit of repeater-less communication, which is known as the PLOB bound \cite{PLOB17}.

In recent decades, research on CVQKD has evolved rapidly.
The Gaussian-modulated CVQKD protocol employing coherent state \cite{GG02} has been theoretically proved to be secure under arbitrary collective attack \cite{COL06} and coherent attacks \cite{COR09}, even taking finite-size regime into account \cite{FIN13,FIN17}. What's more, its composable security has been fully proven \cite{COM15}.
Furthermore, this protocol has been experimentally proved to be feasible both in laboratory \cite{GG03,LAB13} and field tests \cite{FIE16}.
A recent demonstration of all-fiber Gaussian-modulated CVQKD protocol has extended the secure transmission distance of CVQKD over 100 km in the laboratory \cite{150KM}, which showing its potential of being an appealing solution for metropolitan quantum networks.

Theoretically, the security analysis of CVQKD relies on some ideal assumptions, which is hard to satisfy in the experimental implementation \cite{IMP00,IMP08,IMP10}.
The imperfect devices lead to the practical security loopholes, which will be utilized by eavesdroppers to take quantum attack strategies, such as local oscillator (LO) fluctuation attack \cite{LOF13}, calibration attack \cite{CAL13}, LO wavelength attack \cite{WAV13} and detector  saturation attack \cite{SAT16}. Most of these attack strategies mainly focus on the loopholes of the practical imperfect detectors. The attackers employ these loopholes to reduce the estimated additional noise by manipulating the measurement results of the detector, causing Alice to overestimate the security key rate, which in turn hides the extra noise introduced by the attack process.

In order to eliminate all the security hazard of the detector end effectively, continuous-variable measurement-device-independent (CV-MDI) QKD has been proposed by several groups independently \cite{MXC14,LZY14,ZYC14,STN15}.
In CV-MDI-QKD protocol, two legitimate parties, Alice and Bob, are both senders. An untrusted third party, Charlie, is introduced to receive quantum states sent by Alice and Bob,  performing Bell-State Measurement (BSM) and communicating the measurement result to generate the secure keys.
Since the measurement part of the protocol is performed by an untrusted third party, the security of the protocol no longer counts on the perfect detector. Therefore, CV-MDI-QKD can remove all known or unknown side-channel attacks on detectors, which means higher practical security. In recent years, several significant results have been achieved in the theoretical research field of CV-MDI-QKD \cite{COT15,XZ16,PP17,CLU18,CLU218,CZY18,ZYJ18,MHX18}.

\begin{figure*}[htbp]
\resizebox{15cm}{!}{
\includegraphics{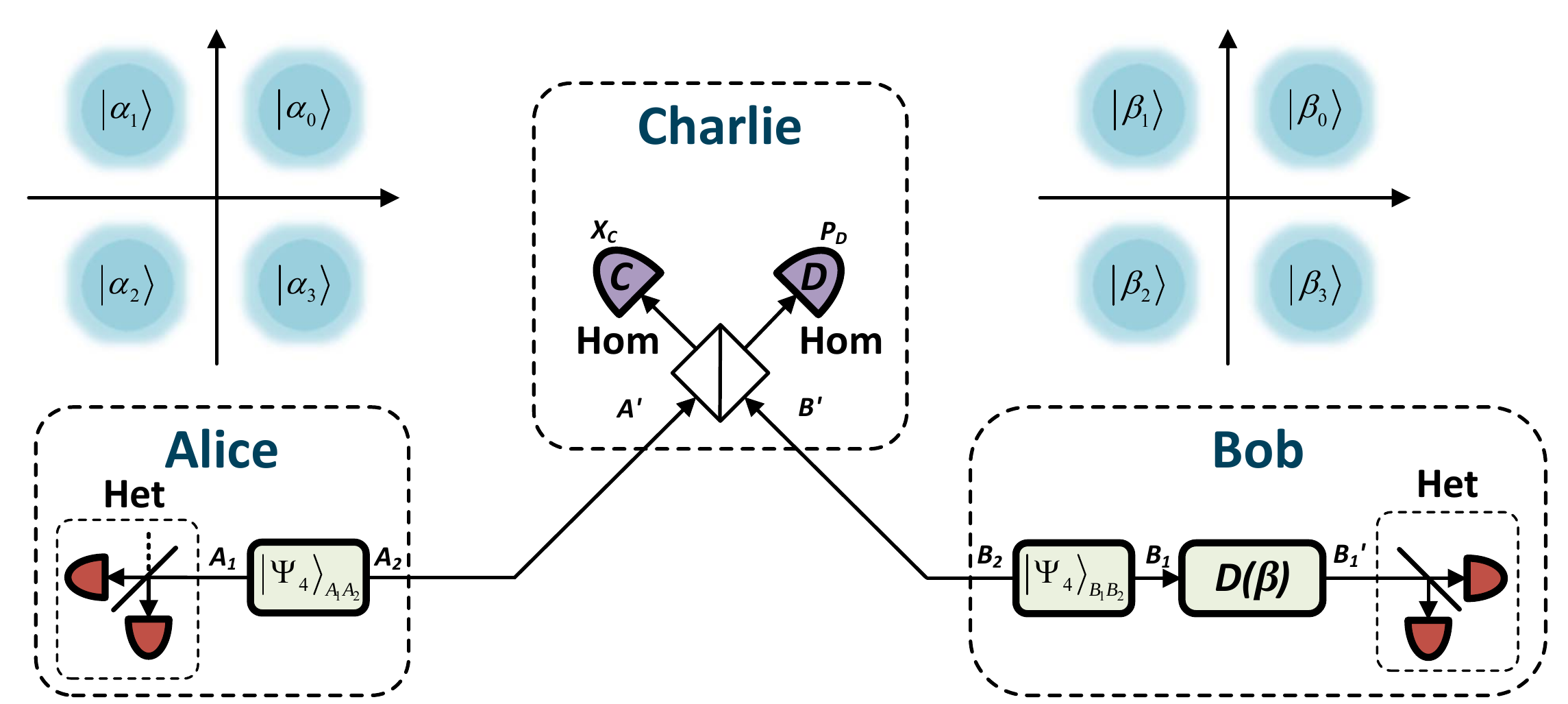}}
\label{fig1}
\caption{(Color online). Schematic diagram of the CV-MDI-QKD protocol with discrete modulation. \({\left| {{\Psi _4}} \right\rangle _{{A_1}{A_2}}}\) and \({\left| {{\Psi _4}} \right\rangle _{{B_1}{B_2}}}\) are all two-mode squeezed states. Het is heterodyne detection. Hom is homodyne detection. \(D(\beta)\) is displacement operation}\label{fig:1}
\end{figure*}

Unfortunately, in practical implementation, the maximal transmission distance of CV-MDI-QKD is unsatisfactory \cite{STN15}. One of the key problems is that the reconciliation efficiency \(\beta\) is quite low for Gaussian-modulated protocols in the case of long distance transmission with low signal-to-noise ratio.
The best error correction codes presently available, such as LDPC codes \cite{LDPC} or turbo codes, work well with discrete (e.g. binary) value under low SNR condition, but have poor performance with continuous (e.g. Gaussian-modulated) value under the same condition.

In order to solve this problem, many efforts are being dedicated. The mainstream approach is to programming an error correction code with higher efficiency under low SNR conditions. This approach is consistent with the way to solve this problem in one-way protocol, such as the code proposed in \cite{CODE}, which can be combined with a multidimensional reconciliation scheme \cite{MULT} and work well in the regime of low SNR. However, this type of error correction codes is designed and implemented with high complexity, and the cost of hardware required is high. What's more, the probability that such codes will succeed in achieving high reconciliation efficiency is quite low \cite{CODE}.

Interestingly, it has been found that the discrete modulation, such as four-state modulation, allows for a much better reconciliation efficiency at low SNR, which lead to longer transmission distance in CVQKD protocol \cite{DMCV09,DECOY}.
In discrete-modulated protocols  \cite{DMCV09,DECOY,FOUR}, the senders prepare a certain number of nonorthogonal coherent states (for four-state scheme, the number is four), and exploit the sign of measured quadratures of each state to encode the bits of secret key rate, while the Gaussian-modulated protocols exploit the quadratures itself to encode information. The sign of measured quadratures is already discrete values, which works well with high efficient error correction codes even at low SNR.
The discrete-modulated protocols are particularly suited in long-distance transmission, whereas they are hard to obtain high key rate at short distance. Surprisingly, a recent work \cite{256} shows the 256 modes discrete-modulated protocol is able to obtain almost the same high key rate with the Gaussian-modulated protocol. Furthermore, discrete modulation is more convenient for experimental implementation and specific operation than Gaussian modulation. Inspired by the aforementioned advantages, we propose a CV-MDI-QKD protocol by employing discrete modulation, which could apparent extent the maximal transmission distance of CV-MDI-QKD protocol without taking additional security vulnerability.

The paper is structured as follows. In Sec.~\ref{PS}, we first introduce the original CV-MDI-QKD protocol, then introduce the model of the CV-MDI-QKD protocol with discrete modulation and the use of decoy states. In Sec.~\ref{Cal}, we derive the secret key of the CV-MDI-QKD protocol with discrete modulation. In Sec.~\ref{PerD},we give the numerical simulation and performance analysis. Conclusion and discussion are drawn in Sec.~\ref{Con}.

\section{CV-MDI-QKD protocol with discrete modulation}\label{PS}  



In this section, we first review the CV-MDI-QKD protocol, especially the entanglement-based (EB) version. Then, we present the CV-MDI-QKD protocol with discrete modulation. In addition, we introduce the use of decoy states for CV-MDI-QKD protocol with discrete modulation, which guarantee the security of this protocol.

\subsection{CV-MDI-QKD Protocol}

Generally, for CV-MDI-QKD protocol, the prepare-and-measure (PM) version is applied to implementation in practice as it is easy to apply, while the equivalent EB version is used for security analysis.
The construction of CV-MDI-QKD protocol is illustrated in Fig.~\ref{fig:1}.
The main steps of EB version can be depicted as follows:

Step 1: Alice and Bob prepare two-mode squeezed states \({\left| {{\Psi _4}} \right\rangle _{{A_1}{A_2}}}\) and \({\left| {{\Psi _4}} \right\rangle _{{B_1}{B_2}}}\), independently. Mode \(A_1\) and \(B_1\) are retained by Alice and Bob respectively, and mode \(A_2\) and \(B_2\) are sent to an untrusted third part Charlie through the quantum channel with length \(L_{AC}\) and \(L_{BC}\), respectively.

Step 2: After receive modes \(A'\) and \(B'\), Charlie interferes the two modes at a beam splitter (BS) and get two output modes \(C\) and \(D\). After the \(x\) quadrature of mode \(C\) and \(p\) quadrature of mode \(D\) are measured by homodyne detections, Charlie announced the measurement results \(\left\{X_C,P_D\right\}\) publicly.

Step 3: Bob uses the received measurement results to modifies mode \(B_1\) through displacement operation \(D(\beta)\) and obtains
\begin{equation}
{\rho_{{{B'}_1}}} = D\left( \beta\right){\rho_{{B_1}}}{D^\dag}\left(\beta\right),
\end{equation}
where \(\rho_{B_1}\) is the density matrix of mod \(B_1\) , \(\beta  = g\left( {{X_C} + i{P_D}} \right)\), and \(g\) is the gain of the displacement. Through heterdyne detection, Bob measures mode \(B'_1\) to get \(\left\{X_B,P_B\right\}\), and Alice measures mode \(A_1\) to get \(\left\{X_A,P_A\right\}\).

Step 4:  Alice and Bob implement parameter estimation, information reconciliation and privacy amplification, finally obtaining a string of secret key.

Through the BSM and Bob's displacement operation, modes modes \(A_1\) and \(B'_1\) are entangled \cite{ENT99}, and \(\left\{X_B,P_B\right\}\) and \(\left\{X_A,P_A\right\}\) are correlated.

In the PM scheme, Alice and Bob prepare coherent states independently and send them to Charlie for BSM. After Charlie announce the measurement results, Bob modifies his data according to the measurement results, while Alice keeps her data unchanged. There is no displacement operation for Bob in the PM scheme. Alice and Bob extract a string of secret key by using parameter estimation, information reconciliation and privacy amplification.

\subsection{Discrete modulation in CV-MDI-QKD}

In CV-MDI-QKD protocol with discrete modulation, Alice and Bob perform discrete modulation operations simultaneously. For the sake of specificity and simplicity, we mainly focus on four-state scheme \cite{DMCV09,FOUR}.
In this scheme, the modulated coherent states consist of four types:
\(\left| {\alpha {e^{i\pi /4}}} \right\rangle \),
\(\left| {\alpha {e^{3i\pi /4}}} \right\rangle \),
\(\left| {\alpha {e^{-3i\pi /4}}} \right\rangle \), and
\(\left| {\alpha {e^{-i\pi /4}}} \right\rangle \),
where \(\alpha\) is a positive number related to the modulation variance of coherent states. The four types are shown in Fig.~\ref{fig:1}.
The modulation variance of coherent state is \({V_M} = 2{\alpha ^2}\).

First, we consider the discrete modulation operation in Alice.
In PM version, the mixture state \(\rho^A_4\), which is send by Alice to Charlie through the quantum channel, is given by
\begin{equation}
{\rho ^A_4} = \frac{1}{4}\sum\limits_{k = 0}^3 {\left| {\alpha _k^4} \right\rangle } \left\langle {\alpha _k^4} \right|.
\end{equation}
In EB version, the  two-mode squeezed state prepares by Alice is \({\left| {{\Psi _4}} \right\rangle _{{A_1}{A_2}}}\), and the variances of the quadrature components of the two modes \(A_1\) and \(A_2\) are all \(V_A\). The two-mode squeezed state prepares by Bob is \({\left| {{\Psi _4}} \right\rangle _{{B_1}{B_2}}}\), and the variances of the quadrature components of the two modes \(B_1\) and \(B_2\) are all \(V_B\), and \(V_A=V_B=1+V_M\).
The modes of \({\left| {{\Psi _4}} \right\rangle _{{A_1}{A_2}}}\) and \({\left| {{\Psi _4}} \right\rangle _{{B_1}{B_2}}}\) is illustrated in Fig.~\ref{fig:1}.the two-mode entangled state prepares by Alice is
\begin{equation}
{\left| {{\Psi _4}} \right\rangle _{{A_1}{A_2}}}  = \sum\limits_{k = 0}^3 {\sqrt {{\lambda _k}} \left| {{\phi^A _k}} \right\rangle } \left| {{\phi^A _k}} \right\rangle = \frac{1}{2}\sum\limits_{k = 0}^3 {\left| {{\psi _k}} \right\rangle_{A_1} } \left| {\alpha _k^4} \right\rangle_{A_2},
\end{equation}
where the non-Gaussian states \({\left| {{\psi _k}} \right\rangle_{A_1} }\) is given by
\begin{equation}
{\left| {{\psi _k}} \right\rangle_{A_1} }  = \frac{1}{2}\sum\limits_{m = 0}^3 {{e^{i(1 + 2k)m\pi /4}}\left| {{\phi^A _m}} \right\rangle },
\end{equation}
with \(m \in \left\{ {0,1,2,3} \right\}\) and
\begin{equation}
\left| {{\phi^A _k}} \right\rangle  = \frac{1}{{{e^{\frac{{{\alpha ^2}}}{2}}}\sqrt {{\lambda _k}} }}\sum\limits_{n = 0}^\infty  {{{\left( { - 1} \right)}^n}\frac{{{\alpha ^{4n + k}}}}{{\sqrt {\left( {4n + k} \right)!} }}\left| {4n + k} \right\rangle },
\end{equation}
for \(k \in \left\{ {0,1,2,3} \right\}\), and
\begin{equation}
\begin{array}{l}
{\lambda_{0,2}} = \frac{1}{{2{e^{{\alpha ^2}}}}}\left[ {\cosh \left( {{\alpha ^2}} \right) \pm \cos \left( {{\alpha ^2}} \right)} \right],\\ \\
{\lambda _{1,3}} = \frac{1}{{2{e^{{\alpha ^2}}}}}\left[ {sinh\left( {{\alpha ^2}} \right) \pm \sin \left( {{\alpha ^2}} \right)} \right].
\end{array}
\end{equation}

As the PM version is equivalent to the EB vision for this protocol, the mixture state \(\rho_4\) can be calculated as
\begin{equation}
{\rho ^A_4} = {\rm{Tr}}\left( {\left| {{\Psi_4}} \right\rangle_{{A_1}{A_2}} \left\langle {{\Psi _4}} \right|_{{A_1}{A_2}}} \right) = \sum\limits_{k = 0}^3 {{\lambda _k}\left| {{\phi^A _k}} \right\rangle } \left\langle {{\phi^A _k}} \right|.
\end{equation}
The covariance matrix \({\gamma _{{A_1}{A_2}}}\) of the bipartite state \({\left| {{\Psi _4}} \right\rangle _{{A_1}{A_2}}}\) can be expressed by
\begin{equation}
{\gamma_{{A_1}{A_2}}} = \left( {\begin{array}{*{20}{c}}
{X{{\rm I}_2}}&{Z_4{\sigma _z}}\\ \\
{Z_4{\sigma _z}}&{Y{{\rm I}_2}}
\end{array}} \right),
\end{equation}
where \({\rm I}_2\) is \(2 \times 2\) identity matrix, \({\sigma _z}=diag(1,-1)\), and
\begin{equation}
\begin{array}{l}
X = \left\langle {{\Psi _4}} \right|1 + a_1^\dag {a_1}\left| {{\Psi _4}} \right\rangle  = 1 + 2{\alpha ^2},\\ \\
Y = \left\langle {{\Psi _4}} \right|1 + a_2^\dag {a_2}\left| {{\Psi _4}} \right\rangle  = 1 + 2{\alpha ^2},\\ \\
{Z_4} = \left\langle {{\Psi _4}} \right|{a_1}{a_2} + a_1^\dag a_2^\dag \left| {{\Psi _4}} \right\rangle  = 2{\alpha ^2}\sum\limits_{k = 0}^3 {\lambda_{k - 1}^{3/2}\lambda _k^{ - 1/2}} .
\end{array}
\end{equation}

Similarly, we set the modulation variance in Bob is still \({V_M} =2{\beta ^2}= 2{\alpha ^2}\), and the two output modes of the bipartite state \({\left| {{\Psi _4}} \right\rangle _{{B_1}{B_2}}}\) are \(B_1\) and \(B_2\).
As Alice and Bob perform the same discrete modulation, the covariance matrix \({\gamma _{{B_1}{B_2}}}\) of \({\left| {{\Psi _4}} \right\rangle _{{B_1}{B_2}}}\) is the same as \({\gamma _{{A_1}{A_2}}}\). Displacement operation does not affect the covariance matrix.

To summarize, in PM version of our protocol, Alice randomly  prepares four nonorthogonal coherent states and send one of them to Charlie, Bob randomly prepares another four nonorthogonal coherent states and send one of them to Charlie. After Charlie performs the BSM on the received two states and announced the measurement results, Alice keeps her sign of quadratures and Bob modifies his sign based on these results. In EB version of our protocol, when Alice performs projection measurement on mode \(A_1\), she only discriminates which coherent state is sent to Charlie, which shows the sign of quadratures of the state. And Bob performs projection measurement on mode \(B'_1\) to get the sign of quadratures, which is associated with the sign obtained by Alice. Both the PM version and EB version use the signs of measured quadratures of states to extract a string of secret key.

\begin{figure}[!h]\center
\resizebox{9cm}{!}{
\includegraphics{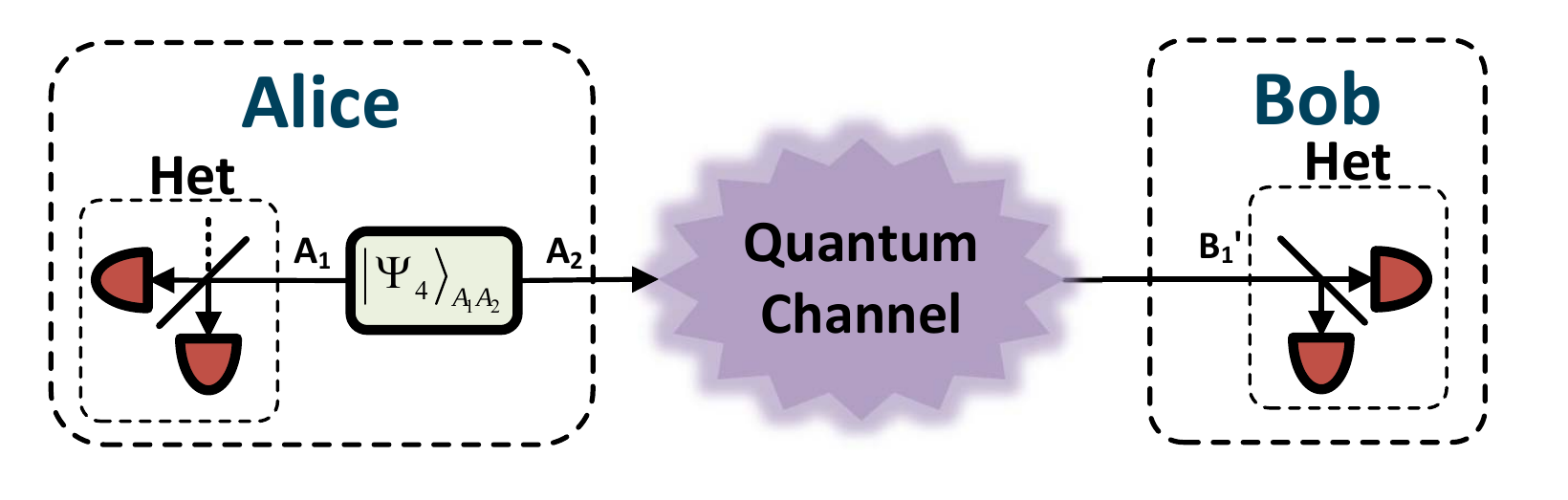}}
\caption{(Color online). Equivalent one-way protocol of the CV-MDI-QKD protocol with discrete modulation in EB version, where Eve is aware of Charlie, channels, the two-mode squeezed state \({\left| {{\Psi _4}} \right\rangle _{{B_1}{B_2}}}\) and the displacement operation \(D(\beta)\).}\label{fig:2}
\end{figure}

\subsection{Decoy states}


In EB version of the our protocol, Alice (Bob) performs the projection measurement, which just makes Alice (or Bob) to discriminate which state is sent to Charlie, but not aware of the quadratures of her (his) quantum mode.
Therefore, the two legitimate users cannot use these experimental data to construct the covariance matrix of the mixed state \(\rho _{AB}\) without the linear channel assumption (LCA) \cite{FOUR}. In other words, LCA is a necessary condition for the security of the protocol.

In order to solve this problem, we introduce decoy states to modify our protocol, which refer to the idea of A. Leverrier et al. \cite{DECOY}. The use of decoy state can be expressed as
\begin{equation}
p\sigma_{key} + (1 - p)\sigma_{decoy} = \sigma _G,
\end{equation}
where \(\sigma _{key} \) is the states used for the key distillation, \(\sigma _{decoy} \) is the decoy states, and \(p\) is a weight between 0 and 1. \(\sigma _G\) is the states satisfying Gaussian distribution and meeting
\begin{equation}
p_{est}\sigma_{est} + (1 - p_{est})\sigma_G = \sigma _{All},
\end{equation}
where \(\sigma_{est}\) is the states used for parameter estimation, \(\sigma _{All}\) are all the states sent through quantum channel. \(\sigma_{est}\) and \(\sigma _{All}\) are all follow Gaussian distribution. \(p_{est}\) denote the fraction of \(\sigma_{est}\).
In practice, Alice randomly prepares \(\sigma^A_{key}\) , \(\sigma^A_{decoy}\) and \(\sigma^A_{est}\) with probability \(p(1-p_{est})\), \((1-p)(1-p_{est})\) and \(p_{est}\) respectively.
The three types of states can be given as
\begin{equation}
\begin{array}{l}
\sigma^A_{key}= \int {{p_{key}}\left( \alpha  \right)} \left| \alpha  \right\rangle \langle \alpha |d\alpha ,\\ \\
\sigma^A_{decoy} = \int {{p_{decoy}}\left( \alpha  \right)} \left| \alpha  \right\rangle \langle \alpha |d\alpha,\\ \\
\sigma^A _{est} = \int {{p_{est}}\left( \alpha  \right)} \left| \alpha  \right\rangle \langle \alpha |d\alpha,
\end{array}
\end{equation}
where the probability distribution \({{p_{key}}\left( \alpha  \right)}\) is chosen as \(p(1-p_{est})\), \({{p_{edcoy}}\left( \alpha  \right)}\) is chosen as\((1-p)(1-p_{est})\) and \({{p_{est}}\left( \alpha  \right)}\) is chosen as \(p_{est}\).
Obviously, the mixed state sent by Alice to Charlie is Gaussian, denoted as \(\sigma^A_{All}\). Bob prepares a Gaussian state \(\sigma^B_{All}\) and sends it to Charlie through quantum channel, taking BSM with \(\sigma^A_{All}\). After these quantum process, Alice tell Bob \(p\), \(p_{est}\) and which state is \(\sigma^A_{key}\), \(\sigma^A_{decoy}\) or \(\sigma^A_{est}\) ,through the classical channel. Then Alice and Bob can establish secure keys through our protocol without LCA. In addition, the use of decoy states can also eliminate the vulnerability exploited by state-discrimination attack \cite{SDAK}, since the discrimination receiver is noneffective for Gaussian states and Eve cannot distinguish whether the state is \(\sigma _{key} \), \(\sigma _{decoy} \) or \(\sigma _{est} \). Otherwise, this method still has certain defects. In practice, these operations will increase the complexity of the system, and the decoy states is hard to be precisely prepared.

\section{Calculation of the secret key rate}\label{Cal}

In this section, we mainly focus on the secret key rate under collective attacks, since they are optimal in the asymptotic limit. The EB scheme of CV-MDI-QKD protocol can be converted into a common one-way CVQKD protocol, where we assume that the two-mode squeezed state \({\left| {{\Psi _4}} \right\rangle _{{B_1}{B_2}}}\) and the displacement operation \(D(\beta)\) are untrusted.
The equivalent one-way protocol is shown in Fig.~\ref{fig:2}.
We denote the secure key rate of the CV-MDI-QKD protocol with discrete modulation and the equivalent one-way protocol are \(K_{DM}\) and \(K_{one}\), respectively.
It is obvious that \(K_{DM} \ge K_{one}\).
Hence, the lower bound of \(K_{DM}\)  can be estimated by using similar covariance matrix of the equivalent one-way protocol.
In PM version of our protocol, the quantum states sent by Alice and Bob are all non-Gaussian states.  According to the optimality of Gaussian attack \cite{GAU}, we can use the method of calculating secret key rate under collective Gaussian attacks to get the low bound of the secret key rate of our protocol under collective attacks. In order to facilitate the analysis, we use the method of calculating \(K_{one}\) under collective Gaussian attacks to obtain \(K_{DM}\).

In addition, we assume the quantum channels from Alice to Charlie and Bob to Charlie are two independent Markovian memoryless Gaussian quantum channels \cite{TWOMODE}. This assumption is based on that the two quantum channels are come from different directions in practical CV-MDI-QKD system, which means that the ambient noise of the two channels should  have weak connection.
Under this assumption, the optimal collective Gaussian attack is taking entangling cloner attacks on each quantum channel independently \cite{COL06,LZY14}. The following calculation and simulation are all based on this attack strategy.

The excess noise and the transmittance of the quantum channel between Alice (Bob) and Charlie are \(\varepsilon_A\) (\(\varepsilon_B\)) and \(\eta_A\) (\(\eta_B\)).
We set both quantum channel losses are \(l\) = 0.2 dB/km,
then the transmittance can be given as
\({\eta_A} = {10^{\frac{{ - l {L_{AC}}}}{{10}}}},{\eta_B} = {10^{\frac{{ - l {L_{BC}}}}{{10}}}}\).
\(\varepsilon\) refers to the equivalent excess noise of the equivalent one-way protocol, which can be calculated as
\begin{equation}
\begin{array}{lll}
\varepsilon  =&1+{\chi _A}+\frac{{{T_B}}}{{{T_A}}}{\left( {{\chi _B} - 1} \right)}
\\ \\
& +\frac{{{\eta_B}}}{{{\eta_A}}}{{\left( {\sqrt {\frac{2}{{{\eta_B}{g^2}}}}\sqrt {{V_B}-1}  - \sqrt {{V_B} + 1} } \right)}^2}
,
\end{array}
\end{equation}
where \({\chi _A} = \frac{1}{{{\eta_A}}} - 1 + {\varepsilon _A}, {\chi _B} = \frac{1}{{{\eta_B}}} - 1 + {\varepsilon _B}\), \(g\) is the gain of the displacement operation in Bob. We adopt \({g^2} = \frac{{2({V_B-1}) }}{{{\eta_B}\left( {{V_B} + 1} \right)}}\) to minimize \(\varepsilon\), then we obtain
\begin{equation}
\begin{array}{lll}
\varepsilon  &= \frac{{{\eta_B}}}{{{\eta_A}}}\left( {{\chi _B} - 1} \right) + 1 + {\chi _A} \\ \\
 &= \frac{{{\eta_B}}}{{{\eta_A}}}\left( {{\varepsilon _B} - 2} \right) + {\varepsilon _A} + \frac{2}{{{T_A}}}.
\end{array}
\end{equation}

\begin{figure}[!h]\center
\resizebox{9cm}{!}{
\includegraphics{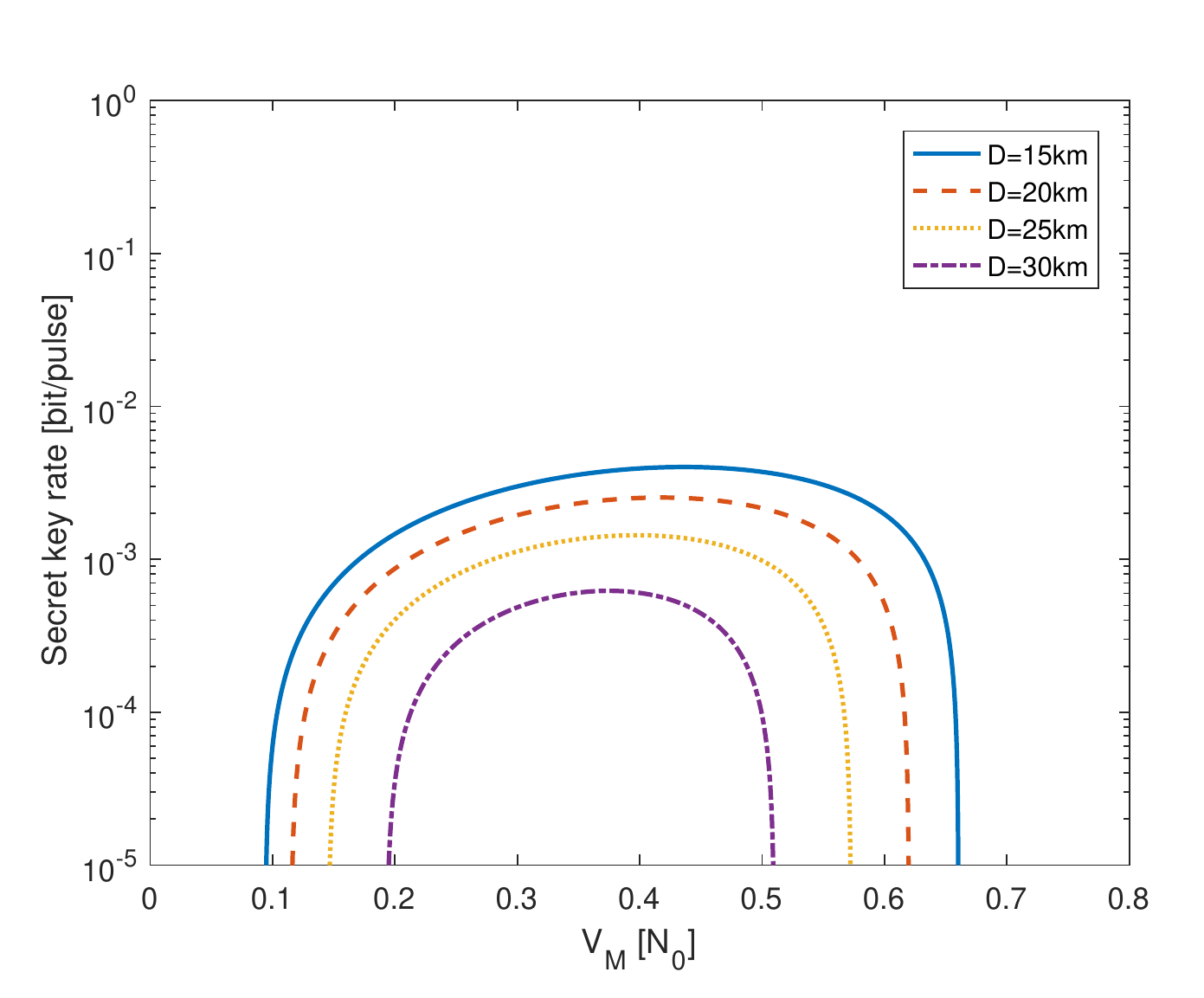}}
\caption{(Color online). Secret key rates as a function of \(V_{M}\) in the extreme asymmetric case. Transmission distances \(D=L_{AC}\) are set to 15 km, 20 km, 25 km and 30 km. \(N_0\) is the shot noise variance. Parameters are fixed as follows: excess noise \(\varepsilon _A=\varepsilon _B=0.002\), reconciliation efficiency \(\beta=90\%\).}\label{fig:3}
\end{figure}
We suppose the homodyne detectors in Charlie are ideal apparatuses, then the total channel-added noise expressed in shot noise units is \(\chi _{t}=\frac{1 }{\eta} -1+ \varepsilon\), where \(\eta = \frac{{{\eta_A}{g^2}}}{2}\) is a normalized parameter which is associated with the quantum channel transmittance of the equivalent one-way protocol \cite{LZY14}.

After the BSM and displacement operation, the covariance matrix of \(\rho _{{A_1}{B'_1}}^{DM}\) is given as

\begin{equation}
\begin{footnotesize}
\begin{array}{l}
\gamma _{{A_1}{{B'}_1}}^{DM} = \left( {\begin{array}{*{20}{c}}
{a{{\rm{I}}_2}}&{c{\sigma _z}}\\
{}&{}\\
{c{\sigma _z}}&{b{{\rm{I}}_2}}
\end{array}} \right)= \left( {\begin{array}{*{20}{c}}
{X{{\rm{I}}_2}}&{\sqrt {\eta} Z_4{\sigma _z}}\\
{}&{}\\
{\sqrt {\eta} Z_4{\sigma _z}}&{{\eta}\left( {Y + {\chi _t}} \right){{\rm{I}}_2}}
\end{array}} \right),
\end{array}
\end{footnotesize}
\end{equation}
where \(X\),\(Y\) and \(Z\) are given in Eq. (9), \({\rm I}_2\) is \(2 \times 2\) identity matrix, \({\sigma _z}=diag(1,-1)\).

The secure key rate of the CV-MDI-QKD protocol with discrete modulation under reverse reconciliation can be calculated as
\begin{equation}
{K_{DM}} = \beta {I^{DM}_{AB}} - {\chi^{DM} _{BE}},
\end{equation}
where \(\beta \in [0,1] \) is the reconciliation efficiency, \(I^{DM}_{AB}\) is the Shannon mutual information between Alice and Bob, \(\chi^{DM} _{BE}\) is the Holevo quantity \cite{HOLEVO} between Bob and Eve. \({I^{DM}_{AB}}\) can be calculated by \cite{KEY}
\begin{equation}
{I^{DM}_{AB}} = 2 \times \frac{1}{2}{\log _2}\frac{{{V_{{A_M}}}}}{{{V_{{A_M}|{B_M}}}}},
\end{equation}
where \({V_{{A_M}|{B_M}}} = {V_{{A_M}}} - \frac{{{c^2}}}{{4{V_{{B_M}}}}}\), \(V_{{A_M}}=(a+1)/2\) , \({V_{{B_M}}}= (b+1)/2\), then
\begin{equation}
I_{AB}^{DM} = {\log _2}\left( {\frac{{a + 1}}{{a + 1 - {c^2}/(b+1)}}} \right).
\end{equation}

The Holevo quantity \({\chi^{DM} _{BE}}\) can be obtained as follows \cite{KEY}
\begin{equation}
\begin{array}{lll}
{\chi^{DM} _{BE}} &= S(E) - S(E|B) \\ \\
&=S(A_1B'_1)-S(A_1|B'_1),
\end{array}
\end{equation}
where \(S(A_1B'_1)\) is a function of the symplectic eigenvalues \(\kappa _{1,2}\) of \(\gamma_{{A_1}{{B'}_1}}^{DM}\), and \(S(A_1|B'_1)\) is a function of the symplectic eigenvalues \(\kappa _3\) of \(\gamma _{{A_1}}^{{m_{{B'_1}}}}\), and
\begin{equation}
\begin{array}{lll}
S(A_1B'_1) =G[(\kappa _1-1)/2]+G[(\kappa _2-1)/2], \\ \\
S(A_1|B'_1)=G[(\kappa _3-1)/2],
\end{array}
\end{equation}
where the Von Neumann entropy
\begin{equation}
G\left( x \right) = \left( {x + 1} \right){\log _2}(x + 1) - x{\log _2}x.
\end{equation}
The symplectic eigenvalues \(\lambda_{1,2}\) can be calculated by
\begin{equation}
\kappa  _{1,2}^2 = \frac{1}{2}\left( {A \pm \sqrt {{A^2} - 4B^2} } \right),
\end{equation}
with the notations
\begin{equation}
\begin{array}{lll}
A=a^2+b^2-2c^2={X^2} + {{\eta}^2}{\left( {Y + {\chi _{t}}} \right)^2} - 2\eta{Z_4^2},\\ \\
B=ab-c^2= \eta\left( {XY + X{\chi _{t}}} - {Z_4^2}\right).
\end{array}
\end{equation}
The covariance matrix \(\gamma _{{A_1}}^{{m_{{B'_1}}}}\) is given by
\begin{equation}
\begin{array}{lll}
\gamma _{{A_1}}^{{m_{{B'_1}}}} &=a{\rm{I}_2}-c{\sigma _z}(b{\rm{I}_2}+{\rm{I}_2})^{-1}c{\sigma _z}
\\ \\
&=[a-c^2/(b+1)]{\rm{I}_2}.
\end{array}
\end{equation}
Then the symplectic eigenvalues \(\kappa _3\) is obtained as
\begin{equation}
\kappa _3=a-c^2/(b+1)=X-{\eta} Z_4^2/[{\eta}(Y+\chi_t)+1].
\end{equation}

\section{Performance analysis}\label{PerD}

In this section, we show the performance of CV-MDI-QKD protocol with discrete modulation compared with the original CV-MDI-QKD protocol.
It has been proven that the configuration with the optimal performance for CV-MDI-QKD protocol is the extreme asymmetric case \cite{STN15}, where Charlie infinitely close to Bob. In our analysis, we are mainly concerned about this case.
In this case, \(L_{BC}=0\), the transmission distance is equal to \(L_{AC}\).
\begin{figure}[!h]\center
\resizebox{9cm}{!}{
\includegraphics{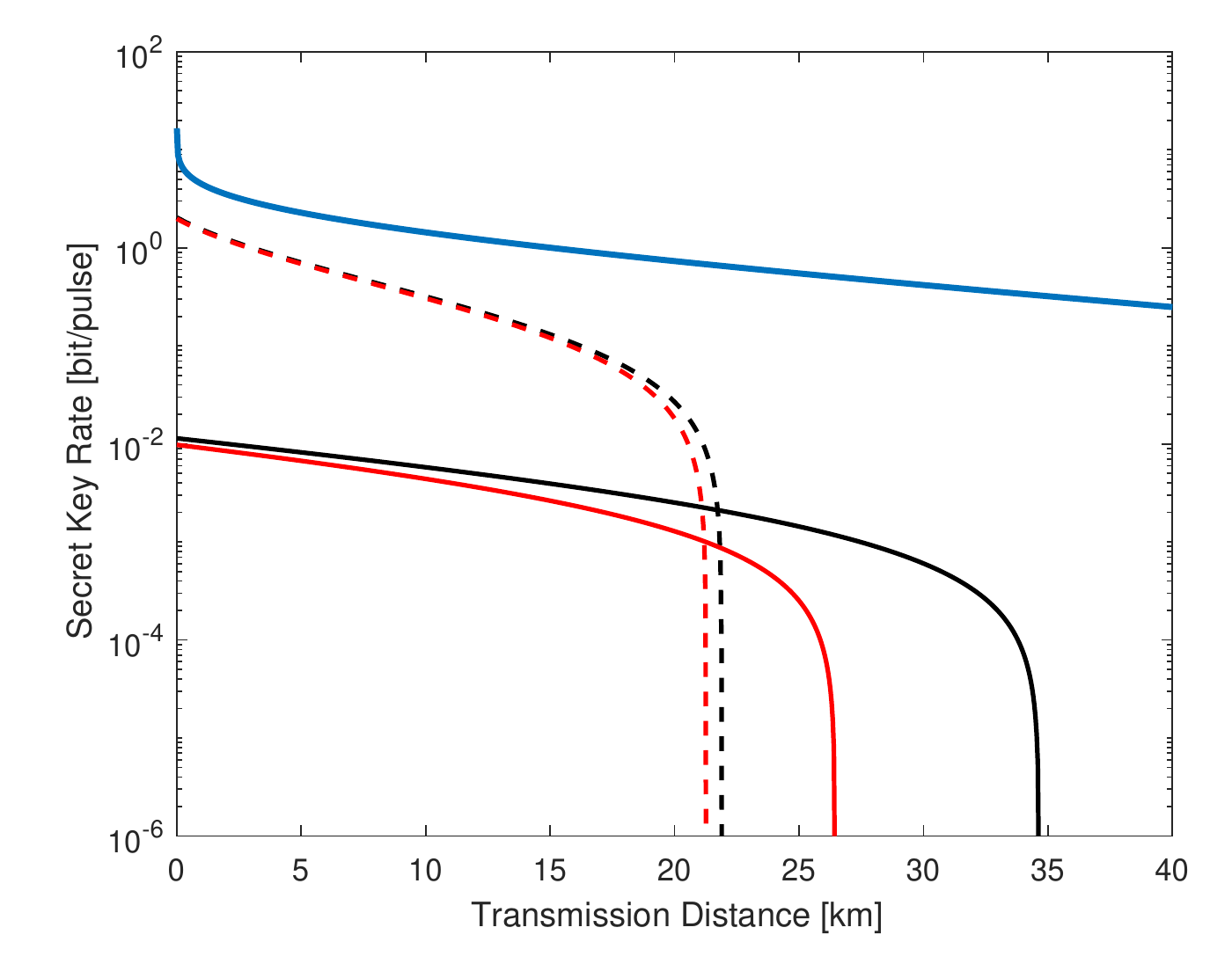}}
\label{fig2}
\caption{(Color online). Secret key rates as a function of the transmission distance in the extreme asymmetric case, where \(L_{BC}=0\). The thin solid lines denote the discrete-modulated CV-MDI-QKD protocol, the dashed lines denote the Gaussian-modulated CV-MDI-QKD protocol. The uppermost heavy solid line denotes the PLOB bound. The various parameters are the excess noises \(\varepsilon _A=\varepsilon _B=0.002\) [upper (black) lines] or \(\varepsilon _A=\varepsilon _B=0.003\) [lower (red) lines]. The modulation variance of Gaussian-modulated protocol and discrete-modulated protocol are 0.4 and 40, respectively. Reconciliation efficiency \(\beta=90\%\).}\label{fig:4}
\end{figure}

The modulation variance \(V_{M}\) is a parameter that will profoundly affect the performance of the prorosed protocol.
In order to extract the optimal secret key rate, we plot the secret key rates as a function of the modulation variance \(V_{M}\) with different transmission distance in extreme asymmetric case, which is shown Fig.~\ref{fig:3}.
It is obvious that the optional areas of \(V_{M}\) are gradually compressed when transmission distance increases, and the secret key rate decreases evidently with the increase of transmission distance. The practicable \(V_{M}\) values of our protocol are quite low than the original CV-MDI-QKD protocol, which means that the quantum signals in our protocol have fewer photons.
Under different transmission distance, the security key rate always have the peak value when \(V_M\) is about 0.4. In other words, the optimal value of \(V_M\) is about 0.4. In the following analysis, we set \(V_M\) of our protocol is 0.4, which leads to the optimal performance.

The plot of Fig.~\ref{fig:4} shows the secret key rates as a function of the transmission distance in the extreme asymmetric case, for both discrete-modulated CV-MDI-QKD protocol and original Gaussian-modulated one, with different excess noise.
The CV-MDI-QKD protocol with discrete modulation is denoted by the thin solid lines, the original CV-MDI-QKD protocol with Gaussian modulation is denoted by the dashed lines, and the PLOB bound \cite{PLOB17}, which gives the ultimate limit of repeater-less quantum communication, is denoted by the uppermost heavy solid line. The modulation variance \(V_M\) of both protocols are all the values used in practical operations, which present the optimal performance of each protocol under practical conditions. \cite{LZY14}.

\begin{figure}[!h]\center
\resizebox{9cm}{!}{
\includegraphics{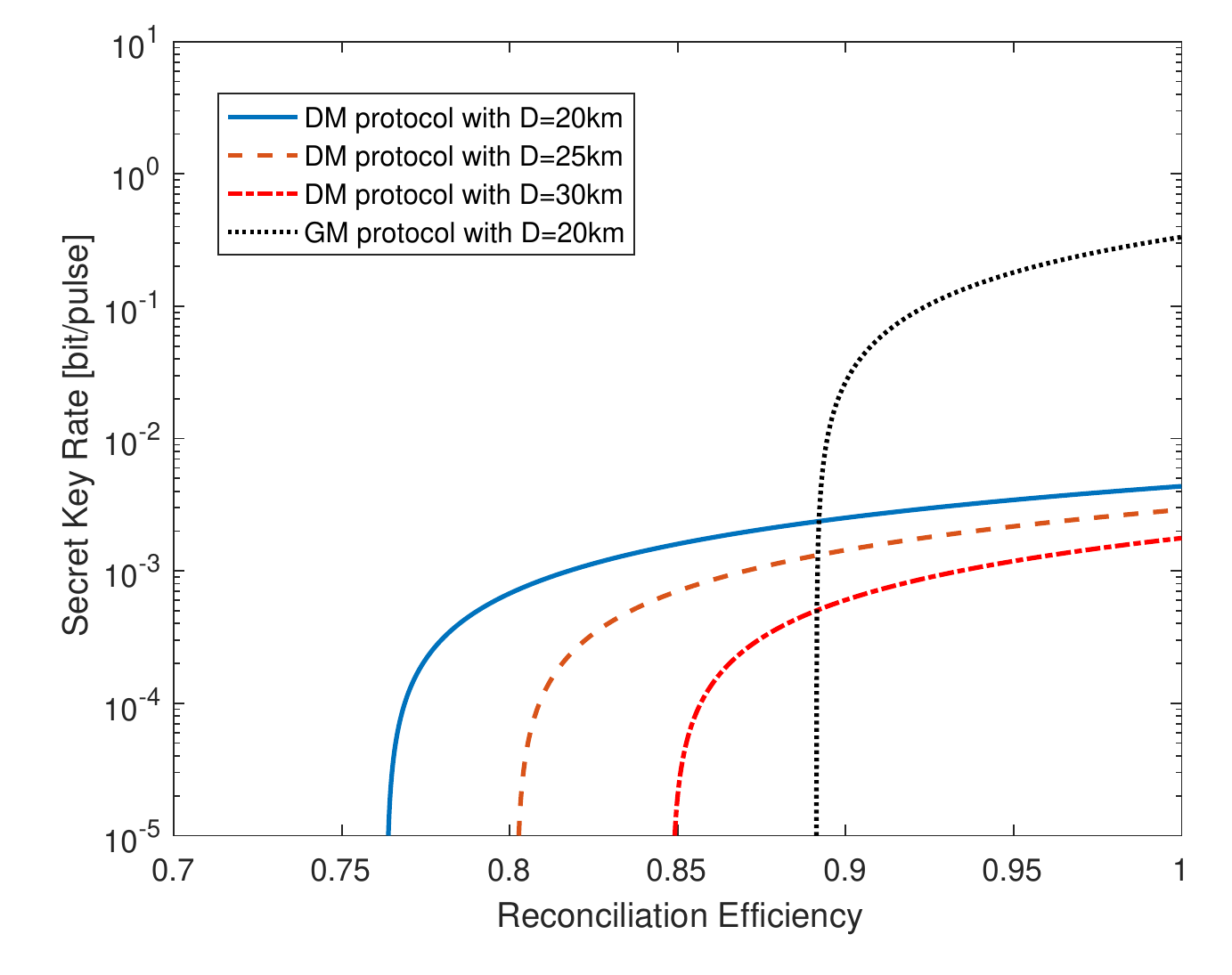}}
\caption{(Color online). Secret key rates as a function of reconciliation efficiency \(\beta\) in the asymmetric case.  Transmission distance denotes as \(D=L_{AC}\). DM protocol denotes the CV-MDI-QKD protocol with discrete modulation, and GM protocol denotes the original CV-MDI-QKD protocol with Gaussian modulation.
Parameters are fixed as follows: excess noise \(\varepsilon _A=\varepsilon _B=0.002\). The modulation variance of Gaussian-modulated protocol and discrete-modulated protocol are 0.4 and 40, respectively.}\label{fig:5}
\end{figure}

It seems clear that the maximum transmission distance of discrete-modulated CV-MDI-QKD protocol is always longer than that of the original Gaussian-modulated one under both excess noises. Compared with the original CV-MDI-QKD protocol, the discrete modulation combined with CV-MDI-QKD protocol can obviously extend the maximum transmission distance, which can effectively expand the scope of application of such protocols. Moreover, when the transmission distance is longer than 21.7 km with the excess noise \(\varepsilon _A=\varepsilon _B=0.002\) (or 21.1 km with the excess noise \(\varepsilon _A=\varepsilon _B=0.003\)), the performance of discrete-modulated CV-MDI-QKD protocol is much closer to the PLOB bound than that of the original Gaussian-modulated CV-MDI-QKD protocol, and the performance gap between the two protocols will become larger and larger with the increase of transmission distance. As the discrete-modulated protocol works well with efficient error correction code in reconciliation process, even in extremely low SNR conditions \cite{DMCV09}, the maximal transmission distance of the discrete-modulated CV-MDI-QKD protocol can be much longer than that of original Gaussian-modulated CV-MDI-QKD protocol in practice.

There is a phenomenon that deserves our attention: within a shorter transmission distance, the performance of the CV-MDI-QKD protocols with discrete modulation is worse than that of original CV-MDI-QKD protocol.
This case is caused by the lower modulation variance of the quantum signal in discrete-modulated CV-MDI-QKD protocol (almost below 1 units of shot noise ), which leads to the secret key rate of discrete-modulated protocol is much lower than the original one in the condition of low-channel-loss.
In addition, as shown in the figure ~\ref{fig:4} , when the excess noise increases, the secret key rate of discrete-modulated CV-MDI-QKD protocol decreases faster than that of the original CV-MDI-QKD protocol. This means that our protocol is more sensitive to the excess noise. The reason for this phenomenon is that the quantum signal intensity and the SNR in discrete-modulated CV-MDI-QKD protocol is much lower than these of original CV-MDI-QKD protocol. However, these defects can be corrected effectively with the participation of noiseless linear amplifier.
Otherwise, Z. Li et al. have proved that discrete-modulated coherent state protocol with 256 modes can reach the high secret key rate almost the same as Gaussian modulation case \cite{256}.


Fig.~\ref{fig:5} depicts a performance comparison of the CV-MDI-QKD protocol with discrete modulation and the original CV-MDI-QKD protocol with Gaussian modulation for different reconciliation efficiency \(\beta\) in the extreme asymmetric case, where \(L_{BC}=0\).
\(V_M\) of both protocols are all the ones provide optimal performance in practice, which is the same as the \(V_M\) in Fig.~\ref{fig:4}.

For the CV-MDI-QKD protocol with discrete modulation, the usable range of \(\beta\) narrows with the increase of transmission distance.
When the transmission distance are 20km, the secret key rate of the CV-MDI-QKD protocols with discrete modulation are lower than that of the original CV-MDI-QKD protocol when the reconciliation efficiency \(\beta\) values in 0.892 to 1.
This is also caused by  the lower quantum signal modulation  variance in discrete modulation. Nevertheless, with \(\beta\) decrease, the secret key rate of the original protocol decrease faster than that of the protocol with discrete modulation. When \(\beta\) is lower than 0.892, the secret key rate of the protocols with discrete modulation get higher than that of the original one. Moreover, the original CV-MDI-QKD protocol will have no secret key when \(\beta\) is lower than 0.889, but the lower bound of \(\beta\) to obtaining secret key for the CV-MDI-QKD protocols with discrete modulation is 0.761. All this means the CV-MDI-QKD protocol with discrete modulation are more tolerant of the reconciliation efficiency. What's more, the protocol with discrete modulation have better combination with efficient error correction code. This will further widen the performance gap between our protocol and the original CV-MDI-QKD protocol over long-distance transmission.


~\\
\section{Conclusion and Discussions}\label{Con} 

In this paper, we have introduced a CV-MDI-QKD protocol with discrete modulation, which can obviously improve the maximal transmission distance. We focus on the analysis of four-state scheme as the representative of discrete modulation scheme. The security of this protocol under arbitrary collective attacks is established with the appropriate use of decoy states.
The discrete-modulated CV-MDI-QKD protocol have commendable
cooperation with efficient reconciliation error correction codes, even in extremely low SNR conditions, which leads to longer secure transmission distance. This property exactly right remedies the defect in the transmission distance of CV-MDI-QKD protocol. The secret key rate simulation results under the asymptotic case shows that the proposed protocol has obvious advantages over the original Gaussian-modulated CV-MDI-QKD protocol in maximal transmission distance, even with the same reconciliation efficiency. What's more, compared with the original Gaussian-modulated CV-MDI-QKD protocol, the experimental implementation of our protocol is more simple and convenient in practice.

In the above analysis, Charlie's homodyne detectors are assumed to be perfect, where the electronic noise \(\upsilon _{el}\) of homodyne detector is 0 and the quantum efficiency \(\eta_{hom}\) of homodyne detector is 1. However,  the practical homodyne detectors in Charlie are not the always ideal apparatuses.
We express the detection-added noise in shot noise units as \({\chi _{hom}} = \left[ {\upsilon _{el}  + \left( {1 - \eta_{hom} } \right)} \right]/\eta_{hom} \). The total noise referred to the channel input with imperfect detectors can be calculated as \({\chi' _{t}} = {\chi _t} + 2{\chi _{hom}}/{\eta_{A}}\). Obviously, the imperfection of the detectors will great increase the total noise. Since the quantum signal intensity of our protocol is much lower than these of the original Gaussian-modulated one, our protocol is more sensitive to the total noise and, especially, the imperfection of the detectors.
But fortunately, through the rational use of noiseless linear amplifier and state-discrimination detection,
we will significantly improve the robustness of our protocol to the imperfections of detectors. Moreover, it would be meaningful to improve the initial secret key rate of the proposed protocol and carry out the experimental implementation in further research.

\begin{acknowledgments}
This work was supported the National key research and development program (Grants No. 2016YFA0302600), the National Basic Research Program of China (Grant No. 2013CB338002) and the National Natural Science Foundation of China (Grants No. 11304397, 61332019, 61505261, 61675235, 61605248, 61671287).
\end{acknowledgments}

\end{document}